%% file: ms.tex
\documentclass[12pt,preprint]{aastex}
\usepackage{natbib}
\bibpunct{(}{)}{;}{a}{}{,}
\newcommand{\hwrange}{1.56 to 1.76 $\mu$m}
\newcommand{\nzfeatures}{33}
\newcommand{\njfeatures}{28}
\newcommand{\nhfeatures}{34}

\shortauthors{Cushing et al.}

\begin{document}

\title{FeH Absorption in the Near-Infrared Spectra of Late M and L Dwarfs}

\author{Michael C. Cushing\altaffilmark{1}}
\affil{Institute for Astronomy, University of Hawai`i, 2680 Woodlawn Drive, 
       Honolulu, HI 96822}
\email{cushing@ifa.hawaii.edu}

\author{John T. Rayner\altaffilmark{1}}
\affil{Institute for Astronomy, University of Hawai`i, 2680 Woodlawn Drive, 
       Honolulu, HI 96822}
\email{rayner@irtf.hawaii.edu}

\author{Sumner P. Davis}
\affil{Department of Physics, University of California, Berkeley, CA 94720}
\email{spdavis@physics.berkeley.edu}

\and

\author{William D. Vacca\altaffilmark{1}}
\affil{Max-Planck-Institut fuer extraterrestrische Physik, Postfach 1312, 
D-85741 Garching, Germany}
\email{vacca@mpe.mpg.de}

\altaffiltext{1}{Visiting Astronomer at the Infrared Telescope Facility,
                 which is operated by the University of Hawai`i under
                 contract from the National Aeronautics and Space
                 Administration.}

\begin{abstract}

	 We present medium-resolution $z$-, $J$-, and $H$-band
spectra of four late-type dwarfs with spectral types ranging from M8 to 
L7.5.  In an attempt to determine the origin of numerous weak absorption 
features throughout their near-infrared spectra, and motivated by the recent 
tentative identification of the $E$ $^4\Pi- A$ $^4\Pi$ system of FeH near 1.6 
$\mu$m in umbral and cool star spectra, we have compared the dwarf spectra to 
a laboratory FeH emission spectrum.  We have identified nearly 100 FeH 
absorption features in the $z$-, $J$-, and $H$-band spectra of the dwarfs.  
In particular, we have identified \nhfeatures\ features which dominate the 
appearance of the $H$-band spectra of the dwarfs and which appear in the 
laboratory FeH spectrum.  Finally, all of the features are either weaker or 
absent in the spectrum of the L7.5 dwarf which is consistent with the 
weakening of the known FeH bandheads in the spectra of the latest L dwarfs.

\end{abstract}


\clearpage

\section{Introduction}

	Absorption bands due to FeH are ubiquitous in the red and 
near-infrared (0.7 $<$ $\lambda$ $<$ 1.3 $\mu$m) spectra of late-type dwarfs.
The most conspicuous FeH feature is the bandhead of the Wing-Ford 
band \citep{wing69} at 0.99 $\mu$m which arises from the 0$-$0 
($\upsilon' - \upsilon''$) transition of the $F$ $^4\Delta- X$ $^4\Delta$ 
system.  \citet{schiavon98} obtained moderate-resolution spectra 
(R$\sim$13,000) of a sample of early to mid M stars to study the 
dependence of the Wing-Ford band on atmospheric parameters.  Their spectra 
show that absorption features due to FeH dominate the spectra of the M 
dwarfs from 0.9850 to 1.0200 $\mu$m.  Other bandheads of this electronic 
system have also been detected in the spectra of late-type dwarfs including 
the 2$-$0 at 0.7786 $\mu$m \citep{tinney98}, the 1$-$0 at 0.8692 $\mu$m 
\citep{kirkpatrick99,martin99}, the 2$-1$ at 0.902 $\mu$m \citep{tinney98}, 
the 0$-$1 at 1.1939 $\mu$m \citep{jones96,mclean00,leggett01, reid01}, and 
the 1$-$2 at 1.2389 $\mu$m \citep{mclean00,leggett01,reid01}.  

	In a recent study of infrared sunspot spectra, 
\citet[hereafter WH]{wallace01} identified 68 lines common to both a sunspot 
spectrum and a laboratory spectrum of FeH between 1.581 and 1.755 $\mu$m. In 
addition, they identified four bandheads in the FeH spectrum at 1.58263, 
1.59118, 1.62457, and 1.61078 $\mu$m.  Based on the theoretical work of 
\citet{langhoff90}, they tentatively assigned this band to the 0$-$0 
transition of the $E$ $^4\Pi -A$ $^4\Pi$ system.

	WH noted that three of the bandheads, at 1.58263, 1.59118, 1.62457 
$\mu$m, could be seen in the low-resolution spectra of late M and L dwarfs 
\citep[e.g.,][]{leggett01}.  These bandheads probably account for the three 
unidentified absorption features in the spectra of L dwarfs at 1.58, 1.613, 
and 1.627 $\mu$m reported by \citet{reid01}.  WH also obtained a 
high-resolution (R=50,000), 70 \AA-wide spectrum of GJ 569B (M8.5 V) centered 
at 1.6578 $\mu$m and identified 13 lines common to both the dwarf spectrum 
and FeH spectrum.  At least in this narrow wavelength range, the dwarf 
spectrum was dominated by FeH absorption lines.

	In the course of conducting a 0.8 to 4.2 $\mu$m spectroscopic survey 
of M, L and T dwarfs, we have identified numerous weak absorption features 
throughout the near-infrared spectra of the M and L dwarfs.  In an attempt to 
determine the origin of these features, and motivated by the work of 
WH, we have compared our spectra to a laboratory FeH emission spectrum to 
determine whether FeH produces these absorption features.

	In this paper, we present medium-resolution $z$-, $J$- and $H$-band 
spectra of four late-type dwarfs with spectral types ranging from M8 to 
L7.5 along with the identification of nearly 100 FeH absorption features.  
In \S2, we discuss the dwarf spectra while in \S3 we discuss the FeH spectrum. 
In \S4 we compare the FeH spectrum and the dwarf spectra and in \S5 we 
discuss our results.  Our conclusions are summarized in \S6.

\section{M and L Dwarf Spectra}

	As part of a infrared spectroscopic survey of $\sim$20 M, L, and T 
dwarfs, three of us (M.C.C., J.T.R., and W.D.V.) have obtained spectra of 
VB 10 (M8 V), 2MASS 1439$+$1929 (L1 V), 2MASS 1507$-$1627 (L5 V), and 
2MASS 0825+2115 (L7.5 V) using SpeX \citetext{Rayner et al., in preparation} 
on the NASA Infrared Telescope Facility.  The resolving power, R, of the 
spectra is $\sim$2000 except for 2MASS 0825+2115 for which R$\sim$1200.  The 
spectral classification of the L dwarfs is from \citet{kirkpatrick99}.  The 
observing strategy and data reduction techniques can be found in Cushing et 
al. (in preparation).  The signal-to-noise of the spectra is $>$50. 

	We use A0 V stars as telluric standards and as a result, care must
be taken in removing the strong hydrogen absorption lines from their spectra.
The standard procedure to remove intrinsic stellar lines is to interpolate 
across them using continuum points on either side of the lines.  Although 
this process works well for isolated lines in regions where the atmospheric 
transmission is flat, it can introduce severe artifacts into the final 
spectra if performed in regions where there are many overlapping stellar 
lines and/or the atmospheric transmission is varying rapidly 
\citep[Figure 3.14]{burgasser02a}.  Our telluric correction method 
\citetext{Vacca et al., in preparation} uses a high-resolution model spectrum 
of Vega to remove the intrinsic spectrum of the A0 V standards.

	An example of the process is shown in Figure 1 which shows the 
$H$-band spectrum of the A0 V standard for VB 10 (lower spectrum), the 
resulting telluric correction spectrum (middle spectrum), and the theoretical
atmospheric transmission for an airmass of 1.5 and 1.6 mm of precipitable 
water vapor as computed with ATRAN \citep{lord92}.  We have also indicated the 
positions of the Brackett lines.  Our telluric correction process removes the 
hydrogen lines in the A0 V star to a level of $<$2\% above the continuum and 
leaves only the atmospheric transmission and instrument response function.  
The telluric correction spectrum is then divided into the dwarf spectra to 
remove telluric absorption and the instrument response function.  The 
telluric correction process does not introduce artifacts into the dwarf 
spectra.

\section{FeH Spectrum}

	One of us (S.P.D.) obtained the laboratory FeH spectrum using a King 
furnace at a temperature of 2700 K and the Fourier Transform Spectrometer at 
the McMath-Pierce Solar Telescope at Kitt Peak.  The furnace sample was 
powdered iron and the filling gas was hydrogen at a pressure of 500 Torr.  
It is possible that water and other contaminants could produce additional 
absorption features in the resulting spectrum.  To avoid this, an enclosed 
optical path was set up using large bore glass tubing that was flushed with 
flowing nitrogen.  The spectrometer was set to a resolution width of 
0.040 cm$^{-1}$ which is equivalent to a resolving power R$\sim$150,000 at 1.6 
$\mu$m.

	A close examination of the spectrum from 0.96 to 2.00 $\mu$m 
verified that the lines analyzed by \citet{phillips87} were present, and 
accounted for all the strong emission lines between 0.78 and 1.34 $\mu$m.  
Unidentified lines outside this region were assumed to be FeH because they 
showed the same line shapes and widths as those of identified lines.  There 
were two regions of unidentified absorption lines in the spectrum, 
interspersed with the emission lines and centered at 1.13 and 1.95 $\mu$m.  
None of the data in this paper covers regions that included absorption lines. 
 
	In the data-reduction process, the continuous background from the 
furnace was subtracted off, and the emission line spectrum set to a baseline 
of zero intensity.  The observed line widths were twice the instrumental 
resolution width, owing somewhat to the high temperature but mostly to the 
high pressure of hydrogen.   In order to compare the FeH spectrum to the dwarf 
spectra, we have degraded the resolution of the FeH spectrum to match that of 
the dwarf spectra by convolving it with a Gaussian of FWHM = 5.4, 6.5, and 
8.1 \AA, the resolution of our observations in the $z$, $J$, and $H$ bands, 
respectively.

\section{Comparison of FeH and Dwarf Spectra}

	In the following sections, we compare $z$-, $J$-, and $H$-band spectra
of the 4 dwarfs to the FeH spectrum.  In addition to lowering the resolution 
of the FeH spectrum to R=2000 (see \S2), we have resampled it onto the 
wavelength grid of the dwarf spectra.  The vacuum wavelengths of the 
FeH features in the $z$, $J$, and $H$ bands are given in Table 1.  Since we 
cannot resolve individual FeH lines with a resolution of $R\sim$2000, all of 
the the FeH features listed in Table 1 are blends of FeH lines.

\subsection{$z$ Band Comparison}

	 Figure 2 shows the $z$-band spectrum of FeH and VB 10 in the lower 
and upper panels, respectively.  The FeH absorption features in this spectral 
region arise primarily from the 0$-$0 band of the $F$ $^4\Delta- X$ 
$^4\Delta$ system.  The strongest feature in the dwarf and FeH spectrum is 
the bandhead of the 0$-$0 band at 0.988 $\mu$m.  We also have identified 
\nzfeatures\ absorption features redward of this bandhead which are present 
in the dwarf spectrum and in the FeH spectrum and have listed them in Table 1.
  
	The feature at 1.0060 $\mu$m arises from the the $F$ 
$^4\Delta_{7/2}- X$ $^4\Delta_{7/2}$ Q-branch \citep{phillips87}.  
\citet{phillips87} also identified another weaker Q-branch of the 
($F$ $^4\Delta_{5/2}- X$ $^4\Delta_{5/2}$) at 0.997904 $\mu$m which we cannot 
unambiguously identify.  There are also three broad features in the VB 10 
spectrum at 0.99513, 1.03490 and 1.06664 $\mu$m which have no counterparts in 
the FeH spectrum.  The feature at 1.03490 $\mu$m is most likely a blend of 
the Fe $a$ $^5$P$_2$ $-$ $z$ $^5$F$_{3}^{\circ}$ line at 1.0343720 $\mu$m, 
the Ca 4$p$ $^1$P$_1$ $-$ 5$s$ $^1$S$_4$ line at 1.0346646 $\mu$m and the two 
weak FeH features at 1.03398 and 1.03554 $\mu$m.  This broad features 
disappears in the spectra of the mid L dwarfs revealing the two weaker FeH 
features.  The other two features at 0.99513 and 1.06664 $\mu$m remain 
unidentified because their shapes and relative strengths do not match the FeH 
features at those wavelength.    

	 Figure 3 shows the $z$-band spectra of the four dwarfs.  The 
\nzfeatures\ features are indicated with dotted lines.  The bandhead, 
Q-branch, and \nzfeatures\ features are present in all of the dwarf spectra 
except for 2MASS 0825$+$2115 (L7.5 V) in which the features are either absent
or considerably weaker.  This dramatic weaking of the 0$-$0 band is 
consistent with the recent finding of \citet{burgasser02a} and will be 
discussed in \S5.

\subsection{$J$ Band Comparison}

	 Figure 4 shows the $J$-band spectrum of FeH and VB 10 in the lower 
and upper panels, respectively.  We believe the FeH features in this spectral 
region arise from the 0$-$1 and 1$-$2 bands of the $F$ $^4\Delta- X$ 
$^4\Delta$ system.  The most prominent features are the bandheads of the 
0$-$1 and 1$-$2 bands at 1.1939 and 1.2389 $\mu$m, respectively.  There are 
also some atomic absorption features seen in the dwarf spectrum namely the 
K I 4$p$ $^2$P $-$ 5$s$ $^2$S multiplet at 1.2432 and 1.2522 $\mu$m, the 
Al I 4$s$ $^2$S $-$ 4$p$ $^2$P multiplet at 1.3117 and 1.3154 $\mu$m and  
the $a$ $^5$P$_{3}$ $-$ $z$ $^5$D$^{\circ}_{4}$ Fe line at 1.1976 $\mu$m.  

	  The features at 1.20907, 1.21126, 1.21348 and 1.22210 $\mu$m were 
previously identified as FeH absorption features by \citet{jones96}.  We have 
identified the feature at 1.22210 $\mu$m as the $F$ $^4\Delta_{7/2} - X$ 
$^4\Delta_{7/2}$ Q-branch \citep{phillips87}.  In addition we have 
identified 24 new features which are present in the dwarf spectrum and 
in the FeH spectrum and have listed them in Table 1 along with the four 
features identified by \citet{jones96}.  The FeH features between 1.2 and 
1.235 $\mu$m arise from the 0$-$1 band while the features longward of 1.24 
$\mu$m arise from the 1$-$2 band although some P-branch lines from the 0$-$1
band are also present.

	 Figure 5 shows the $J$-band spectra of the four dwarfs.  The 
\njfeatures\ features are indicated with dotted lines.  The bandheads, 
Q-branch, and \njfeatures\ features are again present in all of the dwarf 
spectra except for 2MASS 0825$+$2115 (L7.5 V) where the features are either 
weaker or absent.  

\subsection{$H$ Band Comparison}

	 Figure 6 shows the $H$-band spectra of FeH and VB 10 in the lower
and upper panels, respectively.  We believe the absorption features in this 
spectral region are caused by FeH and arise from the 0$-$0 band of the 
$E$ $^4\Pi -A$ $^4\Pi$ system.  Three of the four bandheads identified by 
\citet{wallace01} at 1.58263, 1.59118, 1.62457 $\mu$m are clearly present.  
We have identified \nhfeatures\ features, 32 of which are new, that are 
present in the VB 10 spectrum and in the FeH spectrum and have listed them 
in Table 1.  Two of the feautures at 1.65512 and 1.66100 $\mu$m are blends of 
the 13 FeH lines identified in the spectrum of GJ 569B by \citet{wallace01}.  
There is also a strong absorption feature at 1.64053 $\mu$m which has no 
counterpart in the FeH spectrum and remains unidentified.  This feature is 
located within a much broader absorption feature which does appear in the FeH 
spectrum.

	 Figure 7 shows the $H$-band spectra of the 4 late-type dwarfs along 
with the location of the \nhfeatures\ features described above and three of 
the four bandheads identified by \citet{wallace01}.  The three bandheads and 
\nhfeatures\ features are conspicuous in all of the dwarf spectra except for 
2MASS 0825+2115 (L7.5 V) where the FeH features are considerably weaker.  
This is consistent with the weakening of the FeH absorption features in both 
the $z$- and $J$-band and is further evidence that FeH is the cause of most 
of the absorption seen in the $H$-band spectra of late-type dwarfs.  

\section{Discussion}

	Our results indicate FeH is an important opacity source 
in the atmospheres of late-type dwarfs from 0.99 to 1.7 $\mu$m.  Not only
does FeH produce at least 6 bandheads in this wavelength range, but now 
nearly 100 weaker features which are presumably blends of numerous 
individual absorption lines.  Although \citet{wallace01} found that 
FeH absorption is present in the $H$-band spectra of late-type dwarfs, their 
detections were limited to the three bandheads and 13 lines in a narrow 70 
\AA-wide spectrum.  We have shown that medium-resolution dwarf spectra from 
\hwrange\ consist of numerous absorption features, almost all of which are 
due to FeH. Likewise, the 0.99 to 1.1 $\mu$m dwarf spectra consist of numerous 
absorption features which are also almost all caused by FeH.

	As noted in \S4, the FeH absorption features throughout the 
near-infrared spectra of 2MASS 0825$+$2115 (L7.5 V) are either much weaker 
than the earlier L dwarfs or absent entirely.  \citet{burgasser02a} 
have also found that the Wing-Ford band at 0.99 $\mu$m weakens and then 
disappears in mid to late L dwarfs.  This is consistent with Fe 
condensing out of the atmosphere and forming a cloud layer which drops below 
the photosphere with decreasing effective temperature thus removing 
Fe-bearing molecules (FeH) from the layers above the cloud layer.  However, 
they also find that the bandhead reappears in the early to mid T dwarfs.  
Since this is inconsistent with the scenario just described, 
\citet{burgasser02a} propose holes are present in the cloud layer which allow 
the observer to see to deeper and thus hotter layers where FeH is not 
depleted.  Unfortunately, we cannot test whether the FeH features in the 
$J$- and $H$-band spectra will reappear in early T dwarf spectra since strong 
methane absorption features appear in both the $J$- and $H$-band at these low 
effective temperatures and will mask any FeH features that may be present.

	The appearance of two CrH bandheads of the $A$ $^6\Sigma^+$ $-$ $X$ 
$^6\Sigma^+$ system at 0.86 (0$-$0) and 0.99685 (0$-$1) $\mu$m are a 
defining characteristic of the L spectral type.  In fact, a spectral
index based on the 0$-$0 bandhead is used to spectral type L dwarfs in 
the red-optical \citep{kirkpatrick99}.  We question the identification of the 
0$-$1 bandhead at 0.997 $\mu$m since there are three FeH absorption features 
at this wavelength that are apparent in the spectra of the mid M to 
late L dwarfs.  This may explain the scatter in the spectral index 
defined using the 0$-$1 bandhead \citep{kirkpatrick99}.  Only higher 
resolution spectra will be able to determine to what extent the 0$-$1 CrH 
bandhead is present.

	Finally, it should be noted that the opacities for FeH used in 
current atmospheric models are known to be incorrect (P. Bernath et al., in
preparation).  In addition, an analysis similar to that of \citet{phillips87} 
has yet to be performed on the new $E$ $^4\Pi -A$ $^4\Pi$ system identified by 
\citet{wallace01}.  Atmospheric models are necessary to determine the 
effective temperature scale for late-type dwarfs and until the correct FeH 
opacities are included in these models, obtaining good fits between the 
observations and models near FeH features may be difficult.  Fortunately, 
there is work currently under way to revise the FeH opacities 
(P. Bernath et al., in preparation).  Once this work is completed, it should 
also be possible to constrain the abundance of FeH in the atmospheres of 
late-type dwarfs as has recently be done for CrH \citep{burrows02} and 
CO \citep{noll97}.

\section{Conclusions}

	We have presented $z$-, $J$-, and $H$-band spectra of four dwarfs 
with spectral types ranging from M8 to L7.5.  Using a laboratory
FeH emission spectrum, we have identified nearly 100 FeH absorption features
in the dwarf spectra.  The $H$-band spectra contain \nhfeatures\ features 
which have counterparts in the FeH spectrum.  Taken together with the results 
of \citet{wallace01}, our results suggest that FeH absorption dominates
the spectra of late-type dwarfs from \hwrange.  

\acknowledgments

M. Cushing acknowledges financial support from the NASA Infrared Telescope
facility.   We obtained the atmospheric transmission data from the 
Gemini Observatory, \url{http://www.gemini.edu/sciops/ObsProcess/obsConstraints/ocTransSpectra.html}. We thank Sandy Leggett for useful discussions.  This 
research has made use of the SIMBAD database, operated at CDS, Strasbourg, 
France as well as the Online Brown Dwarf Catalog 
\url{(http://ganymede.nmsu.edu/crom/cat.html)}, which is compiled and 
maintained by Christopher R. Gelino at New Mexico State University.  Finally
we thank the referee, P. C. Stancil, for his useful suggestions.

\clearpage
\begin{figure}
\plotone{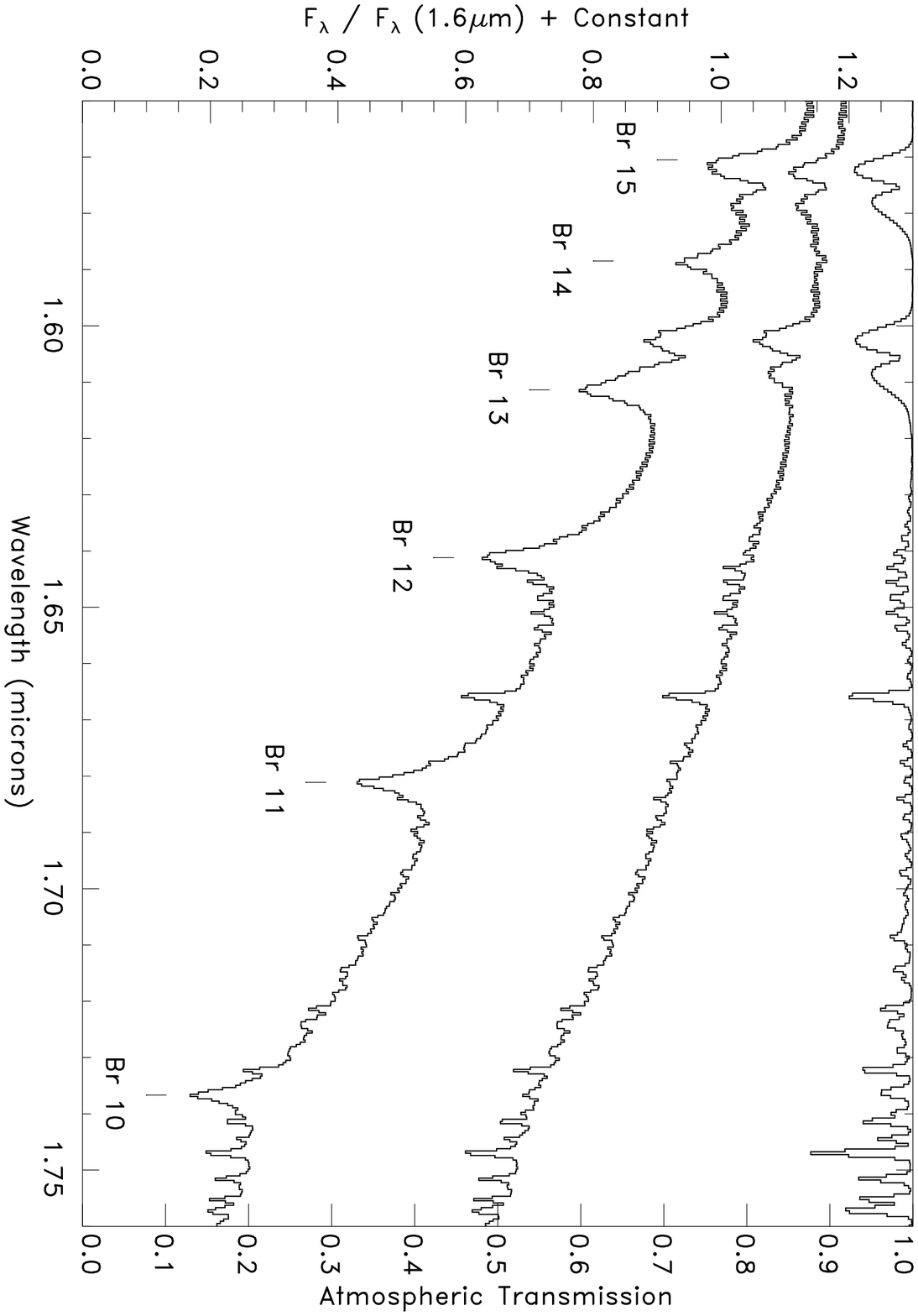}
\vspace{-0.5in}
\caption{The lower spectrum is the observed A0 V standard, the middle 
spectrum is the telluric correction spectrum and the top spectrum is the
theoretical atmospheric transmission.  The positions of the Bracket lines 
are indicated.}
\end{figure}

\clearpage
\begin{figure}
\plotone{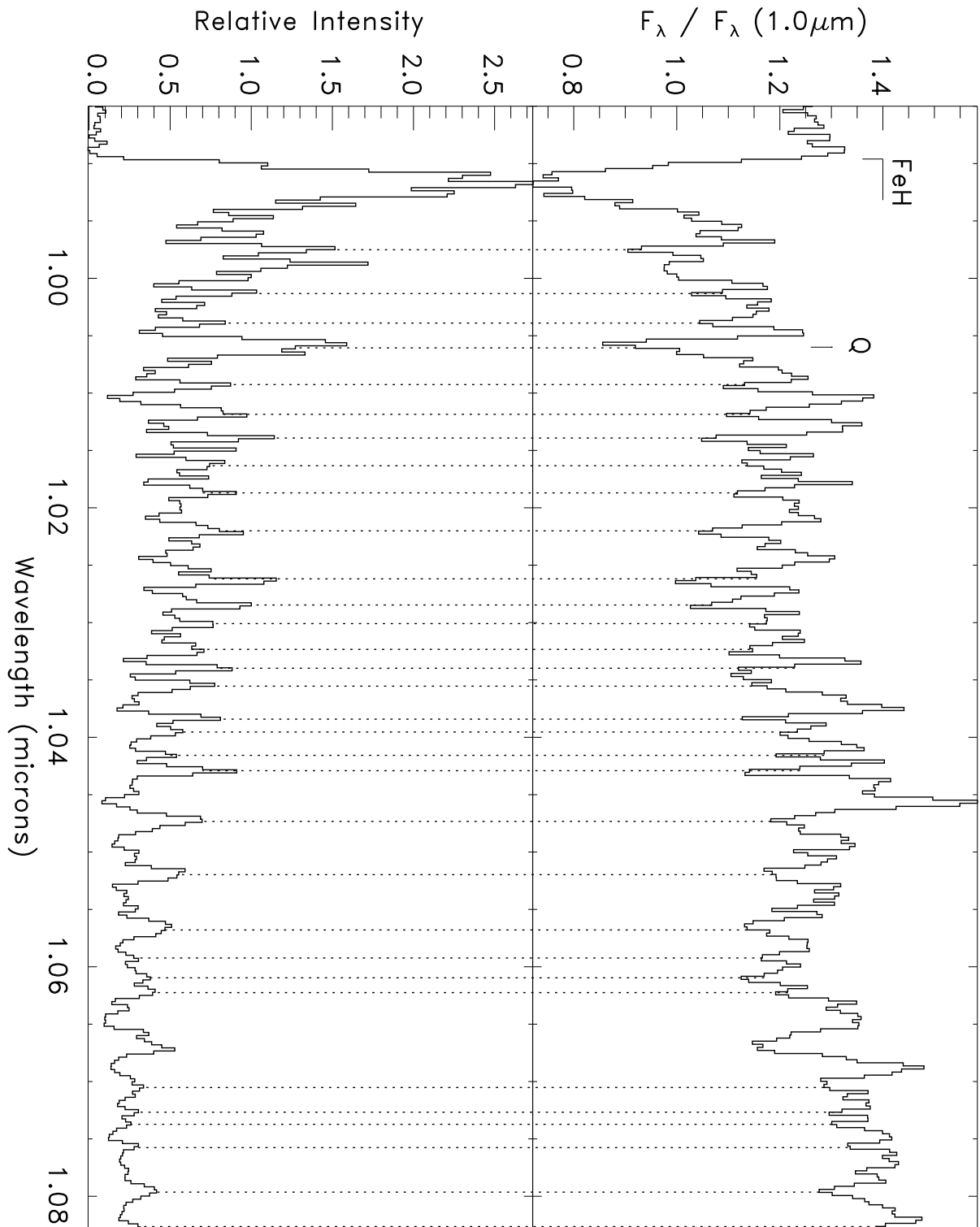}
\vspace{-1in}
\caption{$z$ band:  The lower panel is the FeH spectrum, and the 
upper panel is the VB 10 spectrum.  The location of the 0$-$0 bandhead, 
and Q-branch are indicated.  The \nzfeatures\ features are indicated with 
dotted lines.}
\end{figure}

\clearpage
\begin{figure}
\plotone{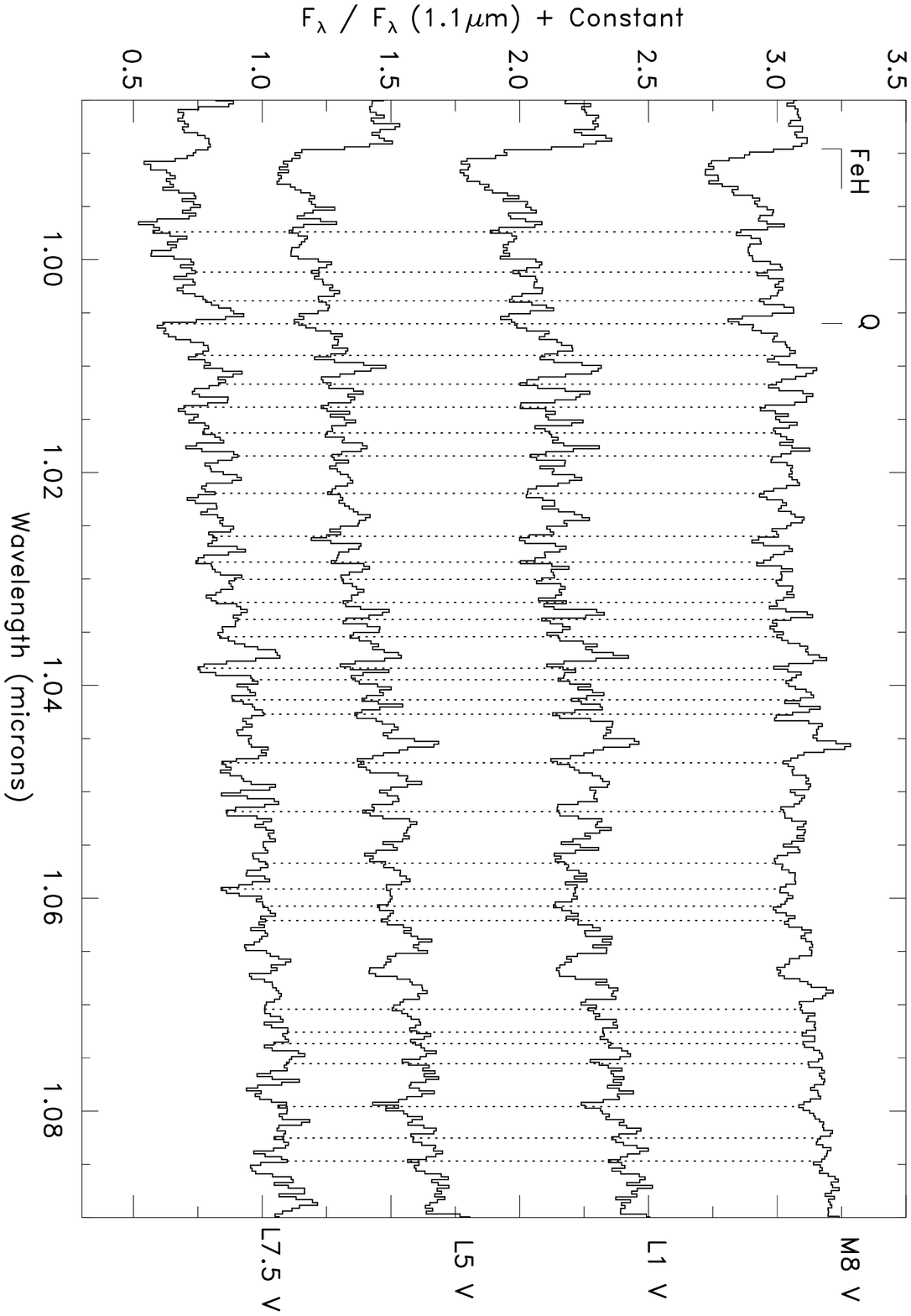}
\vspace{-0.5in}
\caption{$z$ band: The spectra of the 4 dwarfs.  The location of the 
\nzfeatures\ features are shown with dotted lines as well as the bandhead 
and Q-branch feature.}
\end{figure}

\clearpage
\begin{figure}
\plotone{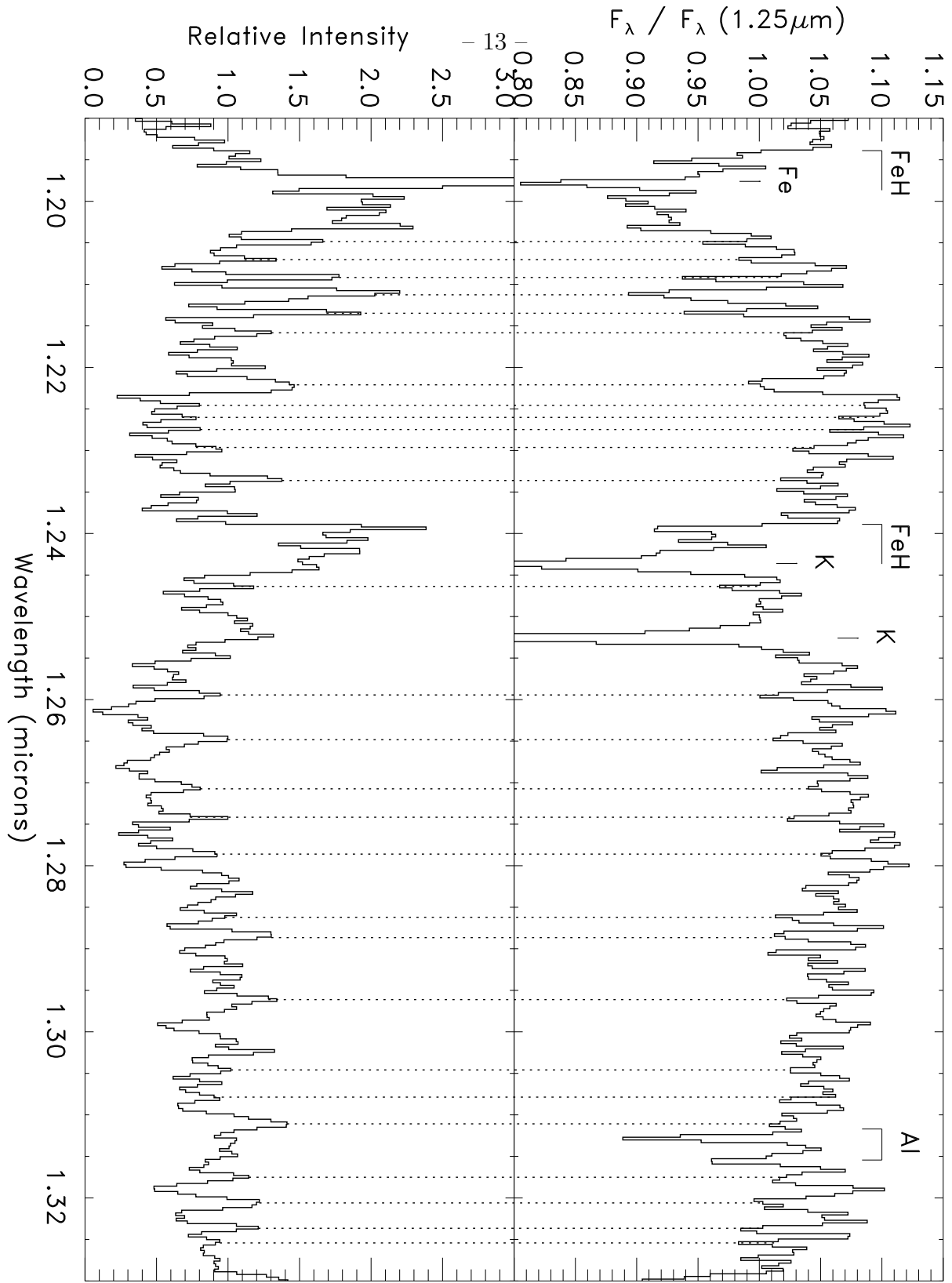}
\vspace{-1in}
\caption{$J$ band:  The lower panel is the FeH spectrum and and the 
upper panel is the VB 10 spectrum.  The location K, Al, and Fe lines, as 
well as the 0$-$1 and 1$-$2 bandheads are indicated.  The \njfeatures\ 
features are indicated with dotted lines.}
\end{figure}

\clearpage
\begin{figure}
\plotone{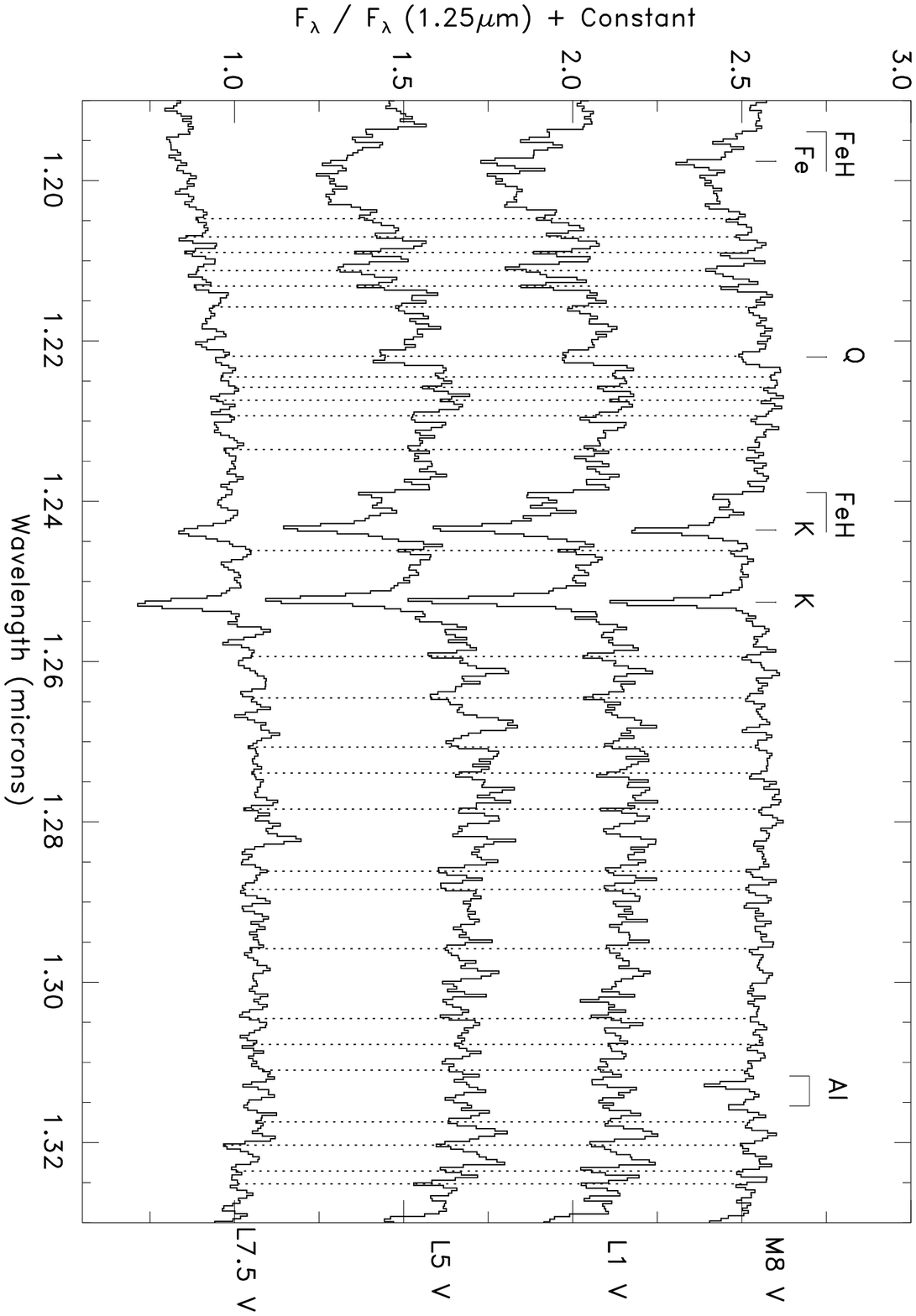}
\vspace{-0.5in}
\caption{$J$ band: The spectra of the 4 dwarfs.  The location of the 
\njfeatures\ features are shown with dotted lines as well as the bandheads 
and Q-branch feature.}
\end{figure}

\clearpage
\begin{figure}
\plotone{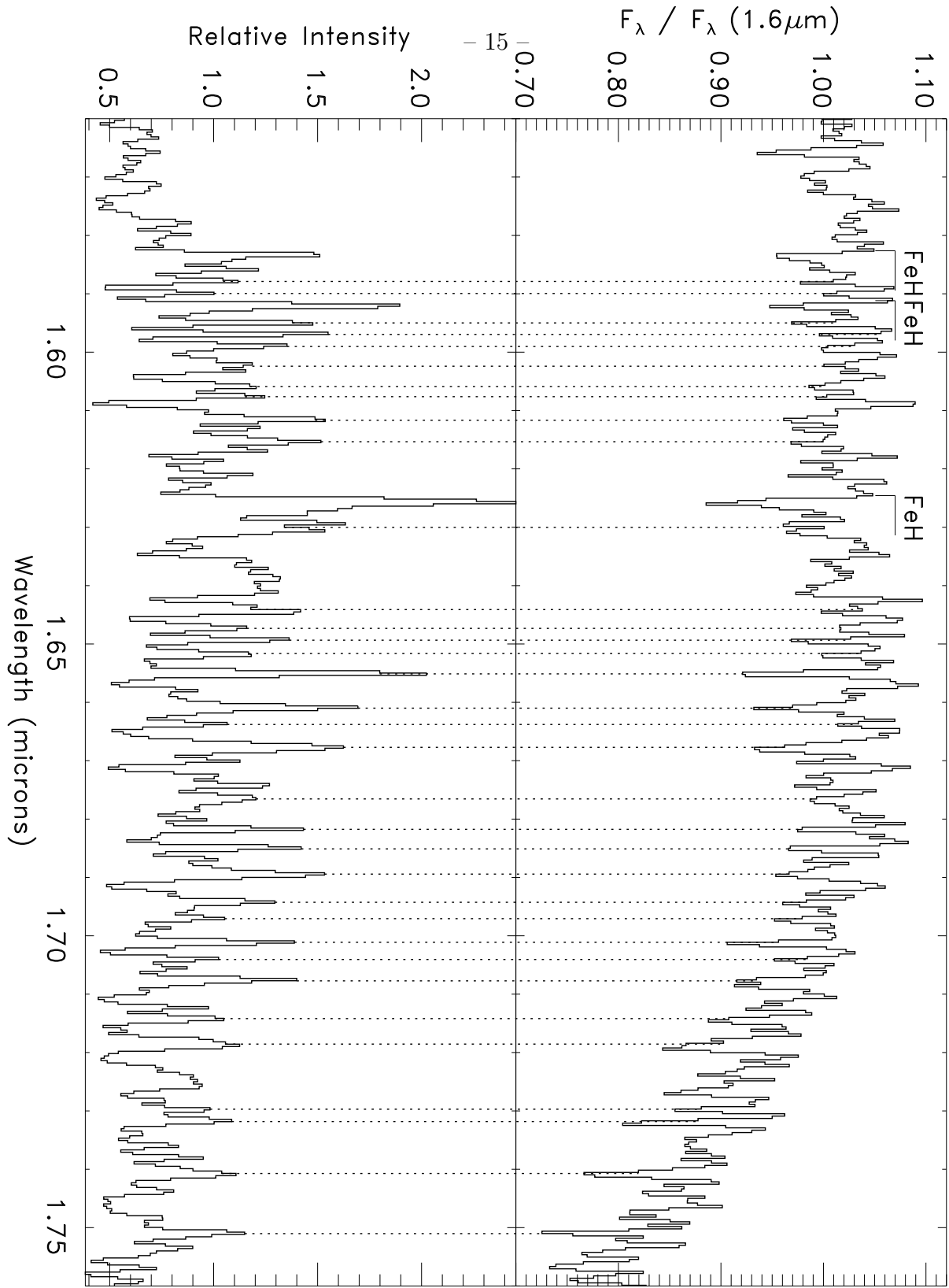}
\vspace{-1in}
\caption{$H$ band: The bottom panel is the FeH emission spectrum and the top 
panel is VB 10 spectrum.  The location of the \nhfeatures\ features are 
indicated with dotted lines.}
\end{figure}

\clearpage
\begin{figure}
\plotone{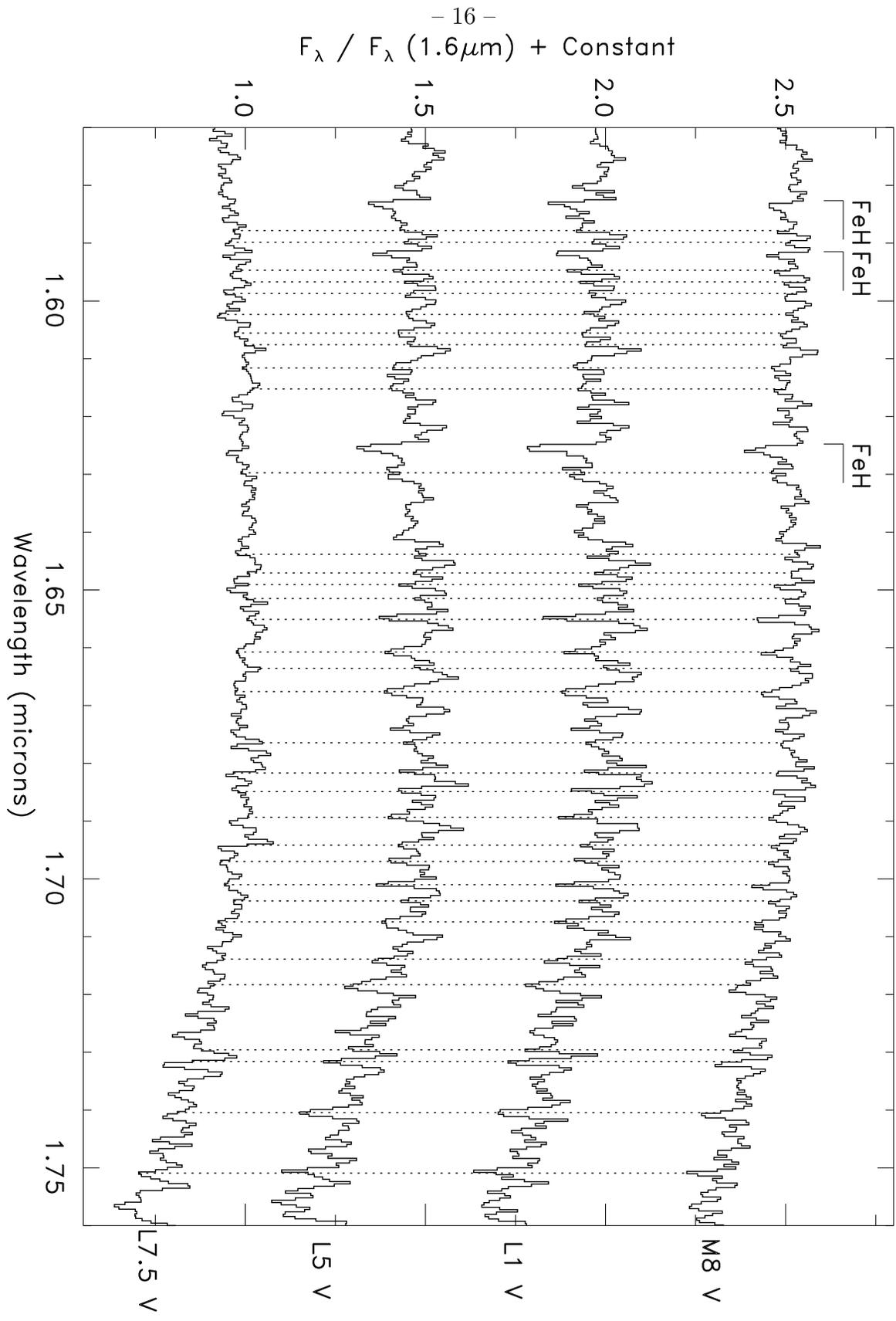}
\caption{$H$ band: The \hwrange\ spectra of the 4 dwarfs.  The location of the 
\nhfeatures\ features are shown with dotted lines.}
\end{figure}

\clearpage

\input{t1.tex}

\end{document}

%% file: t1.tex
% Feature Table

\begin{deluxetable} {cccc} 
\tablewidth{0pt}
\tablenum{1}
\tablecaption{Features Common to FeH and Dwarf Spectra}
\tablehead{\colhead{Vacuum} & \colhead{Vacuum} & 
	   \colhead{Vacuum} & \colhead{Vacuum} \\
          \colhead{Wavelength} & \colhead{Wavelength} & 
	   \colhead{Wavelength} & \colhead{Wavelength} \\
          \colhead{($\mu$m)} & \colhead{($\mu$m)} & 
          \colhead{($\mu$m)} & \colhead{($\mu$m)}} 
\startdata

0.997510                 & 1.06096                  & 1.27074 & 1.64411 \\
1.00133                  & 1.06225                  & 1.27416 &	1.64733 \\
1.00390                  & 1.07052                  & 1.27860 &	1.64935 \\
1.00606\tablenotemark{a} & 1.07267                  & 1.28621 &	1.65164 \\
1.00926                  & 1.07373                  & 1.28866 &	1.65512 \\
1.01185                  & 1.07576                  & 1.29612 &	1.66100 \\
1.01392                  & 1.07964                  & 1.30459 &	1.66379 \\
1.01633                  & 1.08264                  & 1.30786 &	1.66771 \\
1.01870 		 & 1.08492                  & 1.31108 &	1.67655 \\
1.02201 		 & 1.20485                  & 1.31751 &	1.68176 \\
1.02619 		 & 1.20702                  & 1.32060 &	1.68511 \\
1.02847 		 & 1.20917                  & 1.32365 &	1.68943 \\
1.03009 		 & 1.21126                  & 1.32542 &	1.69426 \\
1.03234 		 & 1.21348                  & 1.58788 &	1.69706 \\
1.03398			 & 1.21584                  & 1.58995 &	1.70111 \\
1.03554			 & 1.22210\tablenotemark{a} & 1.59499 &	1.70406 \\
1.03842                  & 1.22458                  & 1.59696 &	1.70772 \\
1.03954 		 & 1.22602                  & 1.59903 &	1.71419 \\
1.04157 		 & 1.22748                  & 1.60241 &	1.71856 \\
1.04291 		 & 1.22964                  & 1.60588 &	1.72971 \\
1.04733			 & 1.23362                  & 1.60763 &	1.73183 \\
1.05197			 & 1.24637                  & 1.61167 &	1.74073 \\
1.05679			 & 1.25944                  & 1.61535 &	1.75103 \\
1.05923                  & 1.26482                  & 1.63002 &         \\
  
\enddata
\tablenotetext{a}{Q-branch}
\end{deluxetable}